\documentclass[aps,prl,twocolumn,superscriptaddress,longbibliography]{revtex4-2}%
\usepackage[breaklinks=true,colorlinks,citecolor=blue,linkcolor=blue,urlcolor=blue]{hyperref}

\usepackage{amssymb,amsmath,amsfonts}
\usepackage{subfigure, latexsym,multirow}
\usepackage[notcite,color,final]{showkeys}
\usepackage{epsfig,graphicx,subfigure,dcolumn,bm}
\usepackage[usenames ,dvipsnames]{xcolor}
\usepackage{slashed}
\usepackage{ulem}
\begin{document}
	\title{Wavefunction-Free Approach for Predicting Nonlinear Responses in Weyl Semimetals} 
	
	\author{Mohammad Yahyavi}
	\affiliation {Division of Physics and Applied Physics, School of Physical and Mathematical Sciences, Nanyang Technological University, 21 Nanyang Link 637371, Singapore}
	\affiliation {Department of Physics, National Cheng Kung University, Tainan, Taiwan}
	
	\author{Ilya Belopolski}
	\affiliation {RIKEN Center for Emergent Matter Science (CEMS), Wako, Saitama 351-0198, Japan}
	
	\author{Yuanjun Jin}
	\affiliation {Division of Physics and Applied Physics, School of Physical and Mathematical Sciences, Nanyang Technological University, 21 Nanyang Link 637371, Singapore}
	
	\author{Yilin Zhao}
	\affiliation {Division of Physics and Applied Physics, School of Physical and Mathematical Sciences, Nanyang Technological University, 21 Nanyang Link 637371, Singapore}
	
	\author{Jinyang Ni}
	\affiliation {Division of Physics and Applied Physics, School of Physical and Mathematical Sciences, Nanyang Technological University, 21 Nanyang Link 637371, Singapore}
	
	\author{Naizhou Wang}
	\affiliation {Division of Physics and Applied Physics, School of Physical and Mathematical Sciences, Nanyang Technological University, 21 Nanyang Link 637371, Singapore}
	
	\author{Yi-Chun Hung}
	\affiliation{Department of Physics, Northeastern University, Boston, MA 02115, USA}
	\affiliation{Quantum Materials and Sensing Institute, Northeastern University, Burlington, MA 01803, USA}
	\affiliation {Institute of Physics, Academia Sinica, Taipei 115229, Taiwan}
	
	\author{Zi-Jia Cheng}
	\affiliation {Laboratory for Topological Quantum Matter and Advanced Spectroscopy (B7), Department of Physics, Princeton University, Princeton, New Jersey 08544, USA}
	
	\author{Tyler A. Cochran}
	\affiliation {Laboratory for Topological Quantum Matter and Advanced Spectroscopy (B7), Department of Physics, Princeton University, Princeton, New Jersey 08544, USA}
	
	\author{Tay-Rong Chang}
	\affiliation {Department of Physics, National Cheng Kung University, Tainan, Taiwan}
	\affiliation {Center for Quantum Frontiers of Research and Technology (QFort), Tainan, Taiwan}
	\affiliation {Physics Division, National Center for Theoretical Sciences, Taipei 10617, Taiwan}
	
	\author{Wei-bo Gao}
	\affiliation {Division of Physics and Applied Physics, School of Physical and Mathematical Sciences, Nanyang Technological University, 21 Nanyang Link 637371, Singapore}
	
	\author{Su-Yang Xu}
	\affiliation {Department of Chemistry and Chemical Biology, Harvard University, Cambridge, MA, USA}
	
	\author{Jia-Xin Yin}
	\affiliation{Department of Physics, Southern University of Science and Technology, Shenzhen, Guangdong 518055, China}
	
	\author{Qiong Ma}
	\affiliation {Department of Physics, Boston College, Chestnut Hill, Massachusetts 02467, USA}

	\author{Md Shafayat Hossain}
	\affiliation{Department of Materials Science and Engineering, University of California, Los Angeles, California 90095, USA}
		\affiliation{California NanoSystems Institute, University of California, Los Angeles, California 90095, USA.}
	
	\author{Arun Bansil}
	\affiliation{Department of Physics, Northeastern University, Boston, MA 02115, USA}
	\affiliation{Quantum Materials and Sensing Institute, Northeastern University, Burlington, MA 01803, USA}
	
	\author{Naoto Nagaosa}
	\affiliation {Fundamental Quantum Science Program, TRIP Headquarters, RIKEN, Wako 351-0198, Japan}
	\affiliation {RIKEN Center for Emergent Matter Science (CEMS), Wako, Saitama 351-0198, Japan}
	
	\author{Guoqing Chang}
	\email{guoqing.chang@ntu.edu.sg}
	\affiliation {Division of Physics and Applied Physics, School of Physical and Mathematical Sciences, Nanyang Technological University, 21 Nanyang Link 637371, Singapore}

	\begin{abstract}
		{By sidestepping the intractable calculations of many-body wavefunctions, density functional theory (DFT) has revolutionized the prediction of ground states of materials. However, predicting nonlinear responses--critical for next-generation quantum devices--still relies heavily on explicit wavefunctions, limiting computational efficiency. In this letter, using the circular photogalvanic effect (CPGE) in Weyl semimetals as a representative example, we realize a 1000-fold computational speedup by eliminating the explicit dependence on wavefunctions. Our approach leverages the one-to-one correspondence between free parameters of Weyl fermions and the associated responses to obtain precise wavefunction-free formulations.  Applying our methodology, we systematically investigated known Weyl semimetals and revealed that Ta$_3$S$_2$ exhibits photocurrents an order of magnitude greater than those observed in TaAs, with potential for an additional order-of-magnitude enhancement under strain. {{To further demonstrate the generality of our approach, we obtained a wavefunction-free formula for the Berry-curvature dipole in Weyl semimetals.}} Our work paves the way for substantially more efficient screening and optimization of nonlinear electromagnetic properties in topological quantum materials. }
	\end{abstract}
	
	\maketitle
Nonlinear electromagnetic responses encompass a rapidly expanding research frontier, linking fundamental science with innovative technologies.  From exotic transport and optical phenomena in quantum materials to the development of advanced electronic chips operating at the cutting edge of hyper-scaling, these nonlinear effects are poised to profoundly shape both our theoretical understanding and the experimental design of next-generation devices~\cite{Nonlinear_Book1,Nonlinear_perspective,Thz1,Nonlinear_Book2,Thz2,Thz3}. Low-frequency nonlinear responses in the terahertz (THz) regime have already established themselves in ultrafast wireless communication systems and high-performance photodetectors~\cite{Thz1,Nonlinear_Book2,Thz2,Thz3}. Further advancements require identifying novel materials with tailored nonlinear electromagnetic properties. Here, (topological) quantum materials offer an especially fertile class of materials ~\cite{QH2,Berry,Hasan2010TI,Zhang2011,Chang2011,Bansil2016, High-throughout1, High-throughout2, High-throughout3, High-throughout4, High-throughout0}, where unique quasiparticles in the momentum space - such as the Weyl and Dirac fermions~\cite{Armitage2018,Nagaosa2020Transport,Hasan2021Weyl,Lv2021Exp} - can exhibit pronounced nonlinear effects driven by their unconventional wavefunction geometries and underlying topologies \cite{NAHE0,NAHE1,Nagaosa2016,NAHE2,Orenstein2021,Ma2021Topology,Bao2022}.  Examples include the presence of substantial nonlinear effects observed in Bi$_2$Te$_3$  \cite{mciver2012control}, WTe$_2$ \cite{WTe,WTe2, H0}, MnBi$_2$Te$_4$ \cite{QM0,QM1},  TaAs~\cite{Ma2017,TaAs_opt,Wu2017Giant}, TaIrTe$_4$ \cite{TaIrTe}, and RhSi \cite{rees2020helicity,ni2021giant}.
	
\begin{figure}[t!] 
		\vspace{0cm}
		\includegraphics[width=\columnwidth]{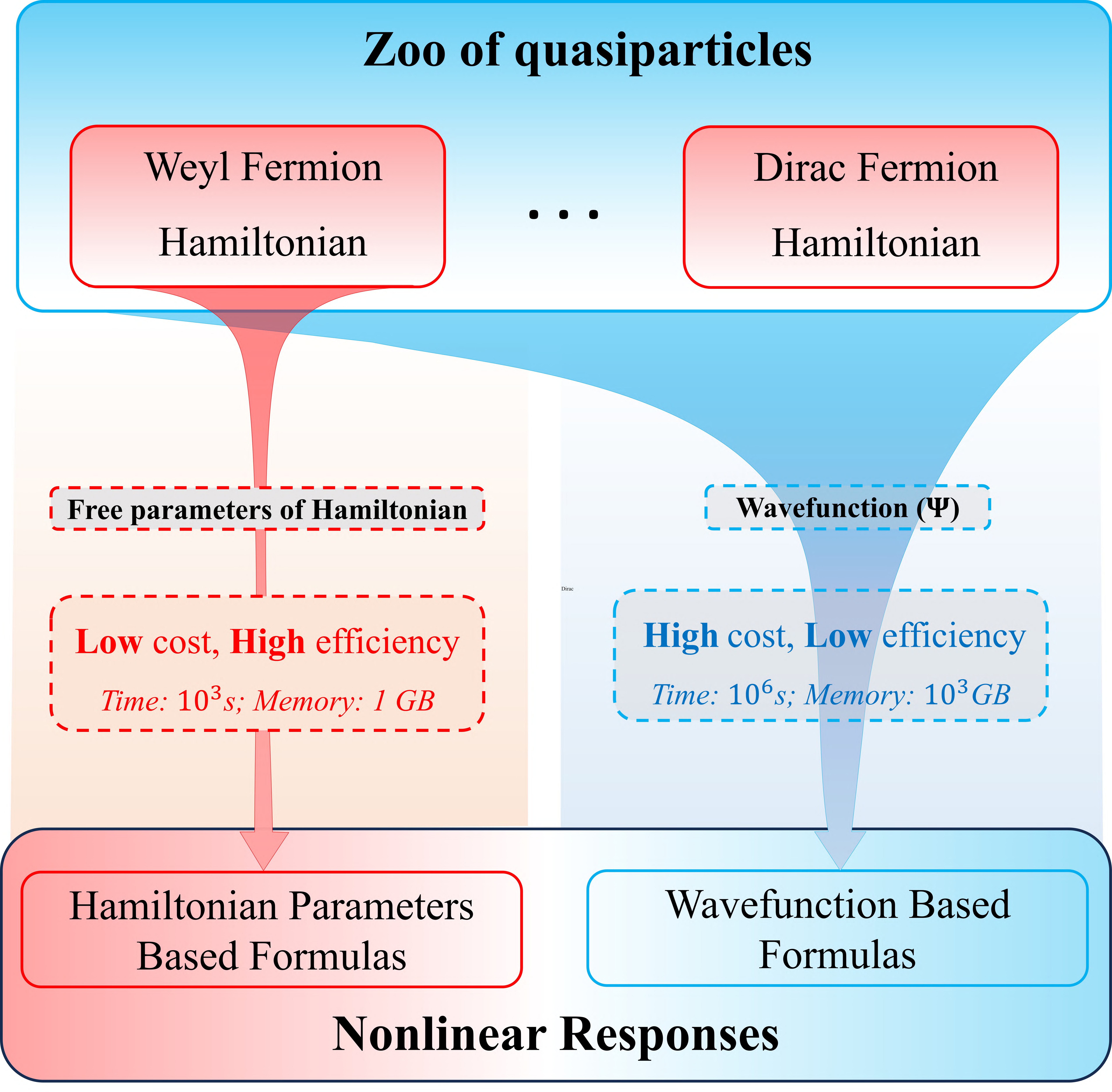}
		\caption{Our work involves obtaining closed-form expressions for nonlinear response tensors directly from Hamiltonian parameters, bypassing expensive wavefunction-based calculations, which makes it possible to greatly reduce computational costs while maintaining good accuracy.}
		\label{Fig.1}
\end{figure}

Despite a growing interest in materials that can host strong nonlinear responses, high-throughput screening of their nonlinear responses remains quite challenging due to the computational complexity associated with integrating intricate wavefunctions across the Brillouin zone (BZ) needed for evaluating these responses \cite{chan2017photocurrents,de2017quantized1,chang2017unconventional,nagaso2020,SC-PRL,DFG-PRB,nagaso2022}. In this letter, we present an approach to circumvent this bottleneck by avoiding explicit wavefunction integrations. Our strategy is to focus on developing wavefunction-free formulas for specific classes of quasiparticles (Fig. \ref{Fig.1}), rather than attempting to handle the entire ``zoo" of quasiparticles simultaneously. For a given type of quasiparticle, once the related Hamiltonian parameters are fixed, the wavefunction geometry is completely determined, which enables us to bypass wavefunction calculations and directly obtain closed-form expressions for nonlinear response tensors from the Hamiltonian parameters alone. Using the circular photogalvanic effect (CPGE) of Weyl fermions as representative examples, we show how one can obtain a simple wavefunction-free empirical formula, which makes it possible to reduce computational costs by roughly three orders of magnitude compared to the wavefunction-based calculations (Fig. \ref{Fig.1}). 

The  second-order tensor of  CPGE-induced injection current  can be expressed as~\cite{chan2017photocurrents,de2017quantized1,chang2017unconventional,nagaso2020,nagaso2022}:
	\begin{equation}\label{CPGE_Org}
		\begin{split}
			\beta_{ab}(\omega) = \frac{\pi e^3}{V\hbar^2} \epsilon_{bcd} \sum_{\textbf{k},n,m} 
			(f^{\textbf{k}}_{n}-f^{\textbf{k}}_{m}) \left( 
			\frac{\partial E^{\textbf{k}}_{m}}{\partial k_a} - 
			\frac{\partial E^{\textbf{k}}_{n}}{\partial k_a} \right) \\
			\times \frac{
				\langle \psi^{\textbf{k}}_{n} | i \frac{\partial \mathcal{H}}{\partial k_c} | \psi^{\textbf{k}}_{m} \rangle
				\langle \psi^{\textbf{k}}_{m} | i \frac{\partial \mathcal{H}}{\partial k_d} | \psi^{\textbf{k}}_{n} \rangle
			}{
				(E^{\textbf{k}}_{n} - E^{\textbf{k}}_{m})^2
			}
			\delta(\hbar \omega - E^{\textbf{k}}_{n} + E^{\textbf{k}}_{m}).
		\end{split}
	\end{equation}
	where $V$ is the volume of the sample, $f^{\textbf{k}}_{n}$ is the Fermi-Dirac distribution, and $| \psi^{\textbf{k}}_{n} \rangle$ denotes the Bloch wavefunction of band $n$. 
	
\begin{figure}[t!] 
		\vspace{0cm}
		\includegraphics[width=\columnwidth]{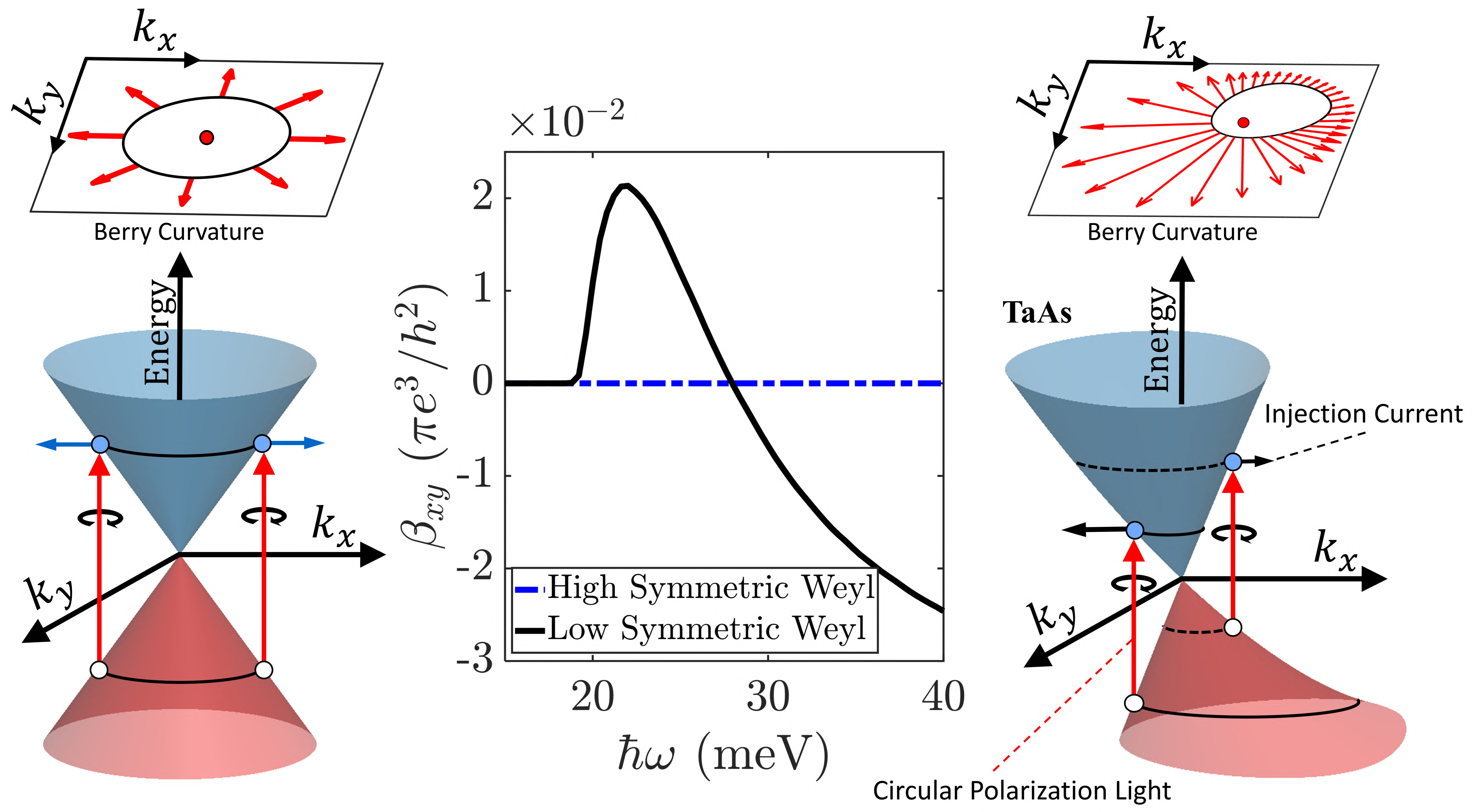}
		\caption{Left: dispersion and Berry curvature of a high symmetry Weyl fermion. Middle: Comparison of the CPGE from a high symmetry Weyl fermion and a Weyl fermion in TaAs. Right: dispersion and Berry curvature of a low symmetry Weyl fermion in TaAs. }
		\label{Fig.2}
	\end{figure}

Using brute-force numerical evaluation of Eq.~\ref{CPGE_Org}, we first consider a Weyl fermion described by the high-symmetry Hamiltonian $\mathcal{H}_0 = \sum_{i} \nu_i \hbar k_i \sigma_i$, where $i = x, y, z$, ${\nu}_i$ represents the velocity, $k_i$ the momentum, and $\sigma_i$ the Pauli matrices~\cite{chan2017photocurrents}. Our calculations show that the transverse CPGE [$\beta_{ab(a \neq b)}$] vanishes [Fig. \ref{Fig.2} middle, blue line]. This is because the band dispersion and Berry curvature distribution are both symmetric around the Weyl node (Fig. \ref{Fig.2}, left), so that contributions from various momentum points cancel out, see Supplemental Material (SM) \cite{SM}  {I}. In many materials, however, low-symmetry conditions induce asymmetry in both the band dispersion and Berry curvature distribution around the Weyl node (Fig. \ref{Fig.2}, right), which breaks the momentum-space balance of photocurrent contributions to yield a net nonzero transverse CPGE (Fig. \ref{Fig.2}, middle panel, black line).

\begin{figure}[t!] 
		\vspace{0cm}
		\includegraphics[width=1\columnwidth]{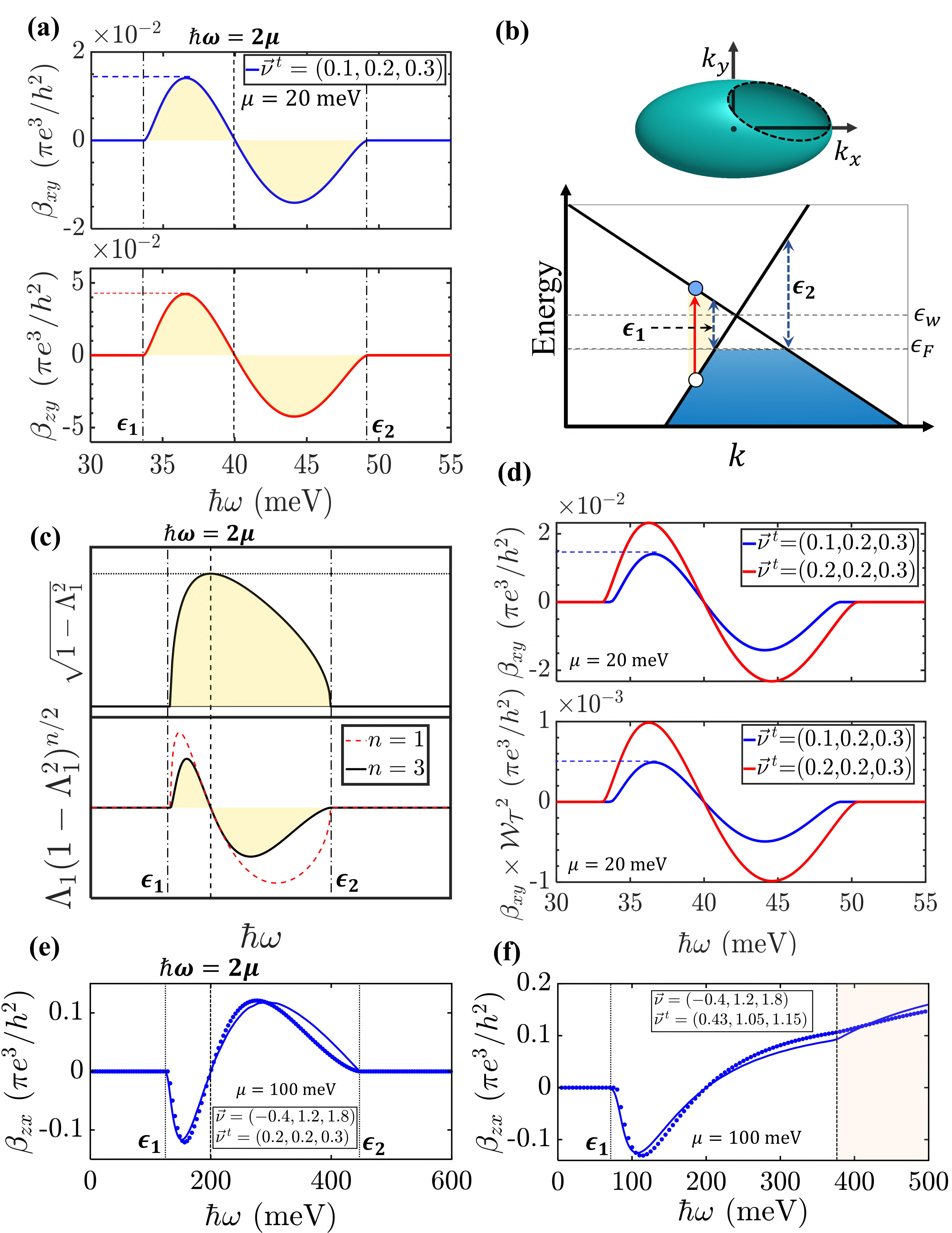}
		\caption{\textbf{(a)} Transverse CPGE of  $\beta_{xy}$ (top) and  $\beta_{zy}$ (bottom) and  for $\vec{\nu}=(2,2,2) $(\AA/s) for $\mu=20$ meV chemical potential.  \textbf{(b)} Schematic of Pauli blocking of a tilted Weyl fermion. The red arrow indicates a representative electronic transition (filled blue circle) in the optical absorption region, highlighted with light coral color. These transitions are such that only a part of the spectrum is Pauli blocked within the photon energy interval $\epsilon_1<\hbar \omega<\epsilon_2$. \textbf{(c)}  Function of $\sqrt{1-\Lambda_1^2}$ (top), $\Lambda_1\sqrt{1-\Lambda_1^2}$, and $\Lambda_1(1-\Lambda_1^2)^{3/2}$ (bottom)  as function of photon energy for $\mu=-50$ meV chemical potential. Here, $\Lambda_1(\hbar\omega)$ is defined as $\frac{1}{\mathcal{W_T}} \left(\frac{2\mu}{\hbar\omega} - 1\right)$. {{\textbf{(d)} Transverse CPGE components \(\beta_{xy}\) (top) and \(\beta_{xy}\times \mathcal{W_T}^2\) (bottom) for \(\vec{\nu} = (2, 2, 2)\) (\text{\AA}/\text{s}) under two different sets of \(\nu_x^t\).}} \textbf{(e)} The transverse CPGE of  a type-I Weyl semimetal model for $\mu=-100$ meV chemical potential. The dotted line corresponds to our proposed formula of Eq. \ref{Eq.tB}, while the solid line shows the results of the numerical calculation from Eq. \ref{CPGE_Org}.  \textbf{(f)} The transverse CPGE of a type-II Weyl semimetal model for a chemical potential of $\mu=-100$ meV are shown. The light orange color indicates the region where  $\hbar \omega \geq  \frac{2\mu}{\mathcal{W_T}-1}$.} 
		\label{Fig.3}
	\end{figure}

To emulate this nonzero CPGE from a Weyl fermion, we introduce a tilting term ($\nu_i^t$) as the first perturbation to the Hamiltonian $\mathcal{H}= \mathcal{H}_0+ \mathcal{H}_{p_1}= \sum_{i} \nu_i \hbar k_i \sigma_i + \sum_{i} \nu_i^t \hbar k_i  \sigma_0 -\mu$, where $\mu$ is the difference between the Weyl node and the Fermi level. Using the conventional Eq.~\ref{CPGE_Org}, we carry out brute-force numerical calculation of $\beta_{ab}$ for a Weyl fermion with anisotropic tilting values $\vec{\nu}^{\,t}=(0.1,0.2,0.3)$ (\AA/s) but isotropic velocities $\vec{\nu}=(0.2,0.2,0.2)$   (\AA/s) [Fig. \ref{Fig.3}(a)]. The results uncover three key features of the transverse CPGE of a titled Weyl fermion. (1) All  \(\beta_{ab(a \neq b)}\)  components are nonzero and oscillate in a trigonometric pattern within a specific energy window. (2) Outside this energy window, \(\beta_{ab(a \neq b)}\) becomes zero. And (3), within the energy window, \(\beta_{ab} = 0\) when \(\hbar \omega = 2\mu\).

In examining optical transitions associated with this Weyl fermion ({SM}~\cite{SM}, Fig. S1), we find an energy range where the electron excitation by photons is asymmetric in momentum space [Fig. \ref{Fig.3}(b)] \cite{nagaso2020}, which explains the presence of the aforementioned energy window. Specifically, when the photon energy $\hbar \omega$ falls within the range $(\epsilon_{1}, \epsilon_{2})$, where $\epsilon_{1,2} = \frac{2\mu}{1 \pm \mathcal{W_T}}$ and $\mathcal{W_T} = \sqrt{(\frac{\nu_x^t}{\nu_x})^2+(\frac{\nu_y^t}{\nu_y})^2+(\frac{\nu_z^t}{\nu_z})^2}$, optical transitions are permitted only on one side of the Weyl cone but forbidden on the other side due to the tilting, resulting in a nonzero transverse CPGE ({SM}~\cite{SM}, Fig. S2).  When $\hbar \omega \leq \epsilon_{1}$, no optical transitions occur, while for $\hbar \omega \geq \epsilon_{2}$, transitions remain symmetric around the Weyl node ({SM}~\cite{SM} {II}), both leading to a zero transverse CPGE.

To capture the three main features discussed above, we first define $\Lambda_1(\hbar\omega)=\frac{1}{\mathcal{W_T}} (\frac{2\mu}{\hbar\omega} -1)$, which gives $\Lambda_1(\epsilon_{1})=1$, $\Lambda_1(2\mu)=0$, and $\Lambda_1(\epsilon_{2})=-1$. Consequently, $\sqrt{1-\Lambda_1^2}$ only yields nonzero real values within the window $(\epsilon_{1}, \epsilon_{2})$, while outside this range ($\hbar \omega \leq \epsilon_{1}$ or $\hbar \omega \geq \epsilon_{2}$)  the real part of $\sqrt{1-\Lambda_1^2}$ is always zero [Fig.~\ref{Fig.3}(c), top-panel]. To ensure that the formula equals zero at $\hbar \omega =2\mu$, we use the function $\Lambda_1 \sqrt{1-\Lambda_1^2}$ [Fig.~\ref{Fig.3}(c), red line in the bottom panel]. To further improve our fit to encompass the peak center and curve skewness, we modified it to $\Lambda_1 (1-\Lambda_1^2)^{3/2}$, which gives the black line in the bottom panel of Fig.~\ref{Fig.3}(c).

Having captured the overall shape of the transverse CPGE curve for tilted Weyl fermions, we now examine how the CPGE amplitude depends on the parameters $(\nu_i^t, \nu_i, \mu)$. In comparing the peak values of different transverse CPGE components, we find that  $\beta_{ab}/\beta_{cb} $ is directly proportional to $\nu^t_a/\nu^t_c$ (Fig.~\ref{Fig.3}(a), {SM}~\cite{SM}, Fig. S3), which can be explained by the relation: $\beta_{ab} \propto \nu^t_a \nu^t_b$. To verify this relationship, we doubled the tilting parameter $v_x^t$ and performed brute-force numerical calculations of $\beta_{ab}$ for this Weyl fermion with modified parameters. Though the amplitude of $\beta_{ab}$ increased, it is not doubled as expected [Fig.~\ref{Fig.3}(d), top]. This suggests the presence of another ``hidden" factor related to the overall tilt of the Weyl fermions that affects all the $\beta_{ab}$ components. A plausible hypothesis is that this hidden parameter is a function of $\mathcal{W_T}$. Accordingly, we carried out further tests and found that $\beta_{ab} \propto \nu^t_a \nu^t_b/\mathcal{W_T}^2$ [Fig.~\ref{Fig.3}(d), bottom]. By changing the Chern number of the Weyl fermions, we observed that $\beta_{ab}$ changes sign ({SM}~\cite{SM}, Fig. S4), showing that the CPGE of Weyl fermions is also proportional to their Chern number  \cite{chan2017photocurrents,de2017quantized1}. Furthermore, our \textit{in silico} simulations, where we varied only the velocities while holding the tilting term fixed, show that the CPGE is also inversely proportional to $\nu_b^2$ ({SM}~\cite{SM}, Fig. S5). Combining all these results, we have: 
	\begin{equation}
		\beta_{ab(a \neq b)}^{p1} =   \frac{3e^3 \mathcal{C} }{\pi h^2}\left[ \frac{\nu_a^t \nu_b^t}{\nu_b^2} \frac{1}{\mathcal{W_T}^2}  \Lambda_1 (1-\Lambda_1^2)^{3/2} \right]\label{Eq.tB1}
	\end{equation} 
	
	
	To validate the robustness of this formula, we randomly generated a set of parameters $(\nu_i^t, \nu_i, \mu)$ for tilted Weyl fermions and independently calculated the associated CPGE using both the conventional wavefunction-based formula and our new wavefunction-free formula. A strong agreement was found between the two results [Fig.~\ref{Fig.3}(e)], demonstrating the efficacy of our new formula for type-I Weyl fermions with $\mathcal{W_T}  <1$.
	
	Turning to type-II Weyl fermions  ($\mathcal{W_T}  > 1$) \cite{soluyanov2015type}, our analysis shows that Eq.~\ref{Eq.tB1}  remains valid at low frequencies but breaks down at high frequencies [Fig.~\ref{Fig.3}(f), solid line]. This is due to the removal of the right boundary defined by $\epsilon_{2} = \frac{2\mu}{1 - \mathcal{W_T}}$ when $\mathcal{W_T}  > 1$. To account for this effect, we slightly modify Eq.~\ref{Eq.tB1} by introducing $\beta_{ab(a \neq b)}^{II} =   \frac{3e^3 \mathcal{C} }{\pi h^2}\left[ \frac{\nu_a^t \nu_b^t}{\nu_b^2} \frac{1}{\mathcal{W_T}^2}  \Lambda_2 (1-\Lambda_2^2)^{3/2} \right]$, where $\Lambda_1(\hbar\omega)=\frac{1}{\mathcal{W_T}} (\frac{2\mu}{\hbar\omega} -1)$ is replaced by  \(\Lambda_2(\hbar\omega) = \frac{1}{\mathcal{W_T}} \left(\frac{2\mu}{\hbar\omega} + 1\right)\). Here, $\beta_{ab(a \neq b)}^{II}$ provides correction at high frequencies  ($\hbar \omega \geq  \frac{2\mu}{\mathcal{W_T}-1}$) for type-II Weyl fermions, but has zero contribution for type-I Weyl fermions. This leads to a unified formula that applies to both type-I  [Fig. \ref{Fig.3}(e)] and type-II [Fig. \ref{Fig.3}(f)] Weyl fermions:
	\begin{equation}
		\beta_{ab(a \neq b)}^{p1} =   \frac{3e^3 \mathcal{C} }{\pi h^2} \left[\frac{\nu_a^t \nu_b^t}{\nu_b^2}   \frac{1}{\mathcal{W_T}^2}    \sum_{n=1}^2 (-1)^{n+1}  \Lambda_n (1-\Lambda_n^2)^{3/2} \right]\label{Eq.tB}
	\end{equation}

 A comparison of the numerical results in Figs. \ref{Fig.2} and \ref{Fig.3} reveals that the CPGE in Weyl fermions is predominantly driven by the tilting term, $\mathcal{H}_{p_1}$. To obtain higher accuracy in describing this CPGE, our approach can be generally applied to other correlations with the Hamiltonian. For instance, using numerical \textit{in silico} simulations ({SM}~\cite{SM} III), we introduced the warping term $\mathcal{H}_{p_2}=\sum_{i} \delta_i  \hbar^2 k_i^2  \sigma_i$ as a perturbation to the Hamiltonian, allowing us to derive a wavefunction-free CPGE formula under this modified framework:
	\begin{equation}\label{BB}
		\beta_{ab(a \neq b)}^{p2} =  \frac{-2e^3  \mathcal{C}}{3\pi h^2} \left[ \frac{\delta_a  \delta_b }{\nu_a\nu_b^3}  (\hbar \omega)^2 \;\Theta(\hbar \omega-2|\mu|)\right] 
	\end{equation}  
	where $\Theta(x)$ denotes the Heaviside step function. 
	
	When both the tilting and warping corrections are considered, the total $\beta_{ab(a \neq b)}$ is simply the sum of the contributions from tilting and warping. The only modification required is to use the lower characteristic energy determined by the tilting term, $\epsilon_1$, in the warping contribution:
	\begin{equation}
		\begin{split}
			\beta_{ab\,(a \neq b)}^{Empir.} = \frac{e^3 \mathcal{C} }{\pi h^2} \Bigg[
			\frac{3 \nu_a^t \nu_b^t}{\nu_b^2 \mathcal{W}_T^2} 
			\sum_{n=1}^2 (-1)^{n+1} \Lambda_n (1 - \Lambda_n^2)^{3/2} \\
			- \frac{2 \delta_a \delta_b}{3 \nu_a \nu_b^3} 
			(\hbar \omega)^2 \; \Theta(\hbar \omega - \epsilon_1)
			\Bigg]
			\label{Eq.GG2}
		\end{split}
	\end{equation}

	\begin{figure}[t!] 
		\includegraphics[width=\columnwidth]{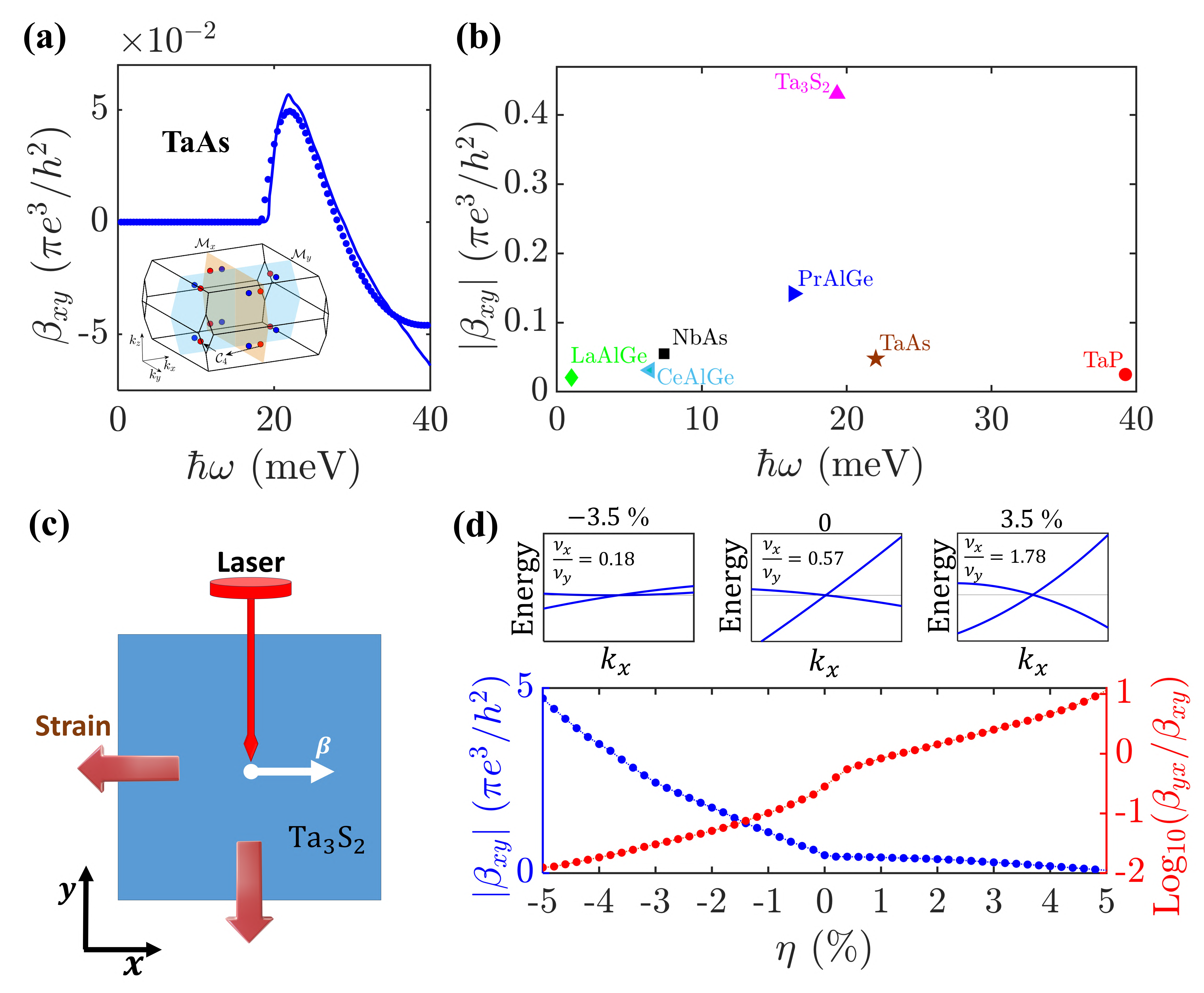}
		\caption{\textbf{(a)} Comparison of $\beta_{xy}$ calculations for sixteen near Fermi level Weyl points (inset) in TaAs with our present approach (dotted line) and conventional formulas (solid line). \textbf{(b)} The maximum frequency-dependent transverse CPGE   $\beta_{xy}$   is compared with the results obtained for low photon energy.  \textbf{(c)} This schematic illustrates the system where a laser is applied to the Ta$_3$S$_2$, with the strain indicated by arrows. \textbf{(d)} Up: Band dispersion of the Weyl node in Ta$_3$S$_2$ under various strain values. Down: Calculated maximum peaks of transverse CPGE (right-blue) $\beta_{xy}$ of Ta$_3$S$_2$ as a function of strain ($\eta$) in the [110] plane.  } 
		\label{Fig.4}
	\end{figure}  

Now we apply our empirical formula to real materials, where there are multiple Weyl fermions in the BZ. While our formula applies to a single Weyl fermion, the CPGE for the other Weyl fermions in the BZ can be efficiently obtained using symmetry relationships ({SM}~\cite{SM} IV).
	
Accurate calculation of the CPGE in a Weyl semimetal using the conventional approach typically involves three steps. (1) Performing first-principles calculations. (2) Generating a real-space tight-binding model (Wannier functions) by fitting the DFT bands. And (3), calculating the CPGE by using the Wannier functions in Eq. \ref{CPGE_Org}. The third step is especially computationally demanding and resource intensive. For example, in our CPGE calculations for TaAs, step 3 required $10^6$ seconds and over 3 TB of CPU memory, while step 1 took only $10^3$ seconds with around 1 GB of CPU memory. In contrast, our wavefunction-free approach streamlines the process to just two steps. (1) Performing first-principles calculations to obtain the material's band structure. And (2), fitting the DFT bands to determine the parameters needed for Eq. \ref{Eq.GG2}. Here, the most time-consuming part is step 1. Our case study of CPGE in TaAs demonstrates that the wavefunction-free method reduces computational costs and time by at least three orders of magnitude while maintaining high accuracy within the low-energy window, where the CPGE is dominated by Weyl fermions [Fig.~\ref{Fig.4}(a)].  
	
By taking advantage of the substantial reduction in computational cost, we studied the CPGE across the well-known Weyl semimetals, including TaAs, TaP, NbP, NbAs, WTe$_2$, MoTe$_2$, WP$_2$, MoP$_2$, LaAlGe,  PrAlGe, CeAlGe, Ta$_3$S$_2$, Co$_3$Sn$_2$S$_2$, SrSi$_2$, HgCr$_2$Se$_4$, CeSbTe, Ta$_2$Se$_8$I, and TaIrTe$_4$ ({SM}~\cite{SM} V and VI) ~\cite{Hasan2021Weyl}. Among these materials, Co$_3$Sn$_2$S$_2$, SrSi$_2$, HgCr$_2$Se$_4$, CeSbTe, and Ta$_2$Se$_8$I exhibit zero transverse CPGE due to symmetry constraints.  Focusing on the low-THz region, where the Weyl semimetals are expected to outperform other quantum materials, we further exclude  NbP, WTe$_2$, MoTe$_2$, WP$_2$, MoP$_2$, and TaIrTe$_4$ because their Weyl fermions are located away from the Fermi level ({SM}~\cite{SM} VI).  Figure \ref{Fig.4}(b) compares the performance of the remaining materials at photon energies below 40 meV, highlighting the frequencies at which the CPGE is maximized. Notably, Ta$_3$S$_2$ demonstrates exceptional performance, exceeding that of TaAs by an order of magnitude [Fig.~\ref{Fig.4}(b)]. 
	
Our wavefunction-free formula also provides a tool for designing experimental setups to enhance CPGE in Weyl semimetals. While it is challenging to manipulate wavefunctions, dispersion-related parameters can be tuned intuitively in experiments. For example, our wavefunction-free formula shows that $\beta_{xy}$ could be enhanced by increasing tilting along ${x}$ and $y$ directions. This can be achieved by applying external strains to the sample. In Ta$_3$S$_2$, we show that a $-5\%$ strain along the [110] plane [Fig.~\ref{Fig.4}(c)] can tune the tilting of Weyl fermions to increase $\beta_{xy}$  by $10$-times of amplitude [Fig.~\ref{Fig.4}(d)]. Eq.~\ref{Eq.tB} also implies $\beta_{yx}/\beta_{xy}=\nu_x^2/\nu_y^2$, so that by tuning the velocity ratio $\nu_x/\nu_y$, CPGE can be enhanced along one direction and suppressed along the other [Fig.~\ref{Fig.4}(d) bottom, red dots], which makes it possible to tune anisoptric of CPGE in Weyl semimetals. 
	
Our approach also makes it possible to identify the conditions under which photocurrents can be maximized ({SM}~\cite{SM} VII). We show that Van der Waals Weyl semimetals in $C_{2v}$ point group could exhibit CPGE values three orders of magnitude greater than TaAs ({SM}~\cite{SM} VIII).  With advances in synthesis techniques, it is now possible to grow Weyl semimetals beyond the existing databases with desired band structures \cite{Ilya2024}. Our work thus provides valuable strategies for the rational design of novel quantum materials capable of achieving much larger photocurrents. 	

To further demonstrate the generality of our approach beyond nonlinear optical effects, we examine the Berry-curvature-dipole (BCD) induced nonlinear Hall effect in Weyl semimetals as a representative case of nonlinear transport phenomena. Analogous to the CPGE, we show that the BCD of Weyl fermions can also be captured through an empirical, wavefunction-free formula that matches well with the conventional approach (see {SM}~\cite{SM} IX.A):
		\begin{equation}\label{NHE01}
	D_{ab(a \neq b)} = -\frac{\text{sign}(\mathcal{C})}{\pi^4}
	\left[
	\frac{\nu_a^t \nu_b^t}{2\nu_b^2}
	+ \frac{2\delta_a \delta_b}{3\nu_a\nu_b^3}\,\mu^2
	+ \frac{11\,\sqrt{\nu_a^{t}\,\nu_b^{t}\,\delta_a\,\delta_b}}{6\,\nu_b^{5/2}\,\nu_a^{1/2}}
	\right].
\end{equation}
Using TaAs as an example, we find that the BCD contribution from trivial pockets is negligible compared to that from Weyl fermions, despite their comparable pocket sizes (see {SM}~\cite{SM} IX.B). This is because the Berry curvature intensity of trivial pockets is far weaker than that of quasiparticles with unconventional quantum geometry. In this sense, our wavefunction-free formulas, which efficiently capture the quasiparticle responses, can be directly employed to estimate the overall nonlinear responses of quantum materials.  Experimental groups can also directly use our approach to rapidly estimate these responses from band structures prior to measurement, thereby identifying promising material candidates.

Beyond enabling efficient computation of nonlinear responses, our approach also makes it possible to provide new physical insights. Conventionally, nonlinear responses are formulated within the {complex expression in terms of Bloch wave functions and their matrix elements.} In contrast, our approach yields a wavefunction-free expression that can be directly interpreted through simple parameters in a Hamiltonian, offering greater physical transparency.  In particular, the inclusion of tilt and warping terms breaks the ``generalized" Lorentz symmetry of Weyl fermions, which intrinsically activates nonlinear responses. Our results for both CPGE and BCD thus demonstrate that the nonlinear responses of Weyl semimetals are fundamentally tied to the breaking of Lorentz symmetry (see {SM}~\cite{SM} X).

In summary, our work establishes a general framework for accelerating the discovery of quantum materials with exotic nonlinear optical and transport responses—an area of broad relevance to condensed matter physics, optics, and materials science. What is more, our framework also provides clear physical insights into these nonlinear responses, unveiling a hidden connection between high-energy physics and condensed matter physics.

\section*{Acknowledgments}
We thank M.~Zahid~Hasan for valuable discussions and insightful comments. Work at Nanyang Technological University was supported by the National Research Foundation, Singapore, under its Fellowship Award (NRF-NRFF13-2021-0010), the Agency for Science, Technology and Research (A*STAR) under its Manufacturing, Trade and Connectivity (MTC) Individual Research Grant (IRG) (Grant No.: M23M6c0100), the Singapore Ministry of Education (MOE) Academic Research Fund Tier 3 grant (MOE-MOET32023-0003), and Singapore Ministry of Education (MOE) AcRF Tier 2 grant (MOE-T2EP50222-0014).  N.N. was supported by JSPS KAKENHI Grant Numbers 24H00197 and 24H02231 and the RIKEN TRIP initiative. The work at Northeastern University was supported by the Air Force Office of Scientific Research under award number FA9550-20-1-0322 and benefited from the resources of Northeastern University’s Advanced Scientific Computation Center, the Discovery Cluster, the Quantum Materials and Sensing Institute, and the Massachusetts Technology Collaborative award MTC-22032. M.Z.H. group acknowledges primary support from the US Department of Energy (DOE), Office of Science, National Quantum Information Science Research Centers, Quantum Science Center (at ORNL) and Princeton University; M.Z.H. acknowledges support from the Gordon and Betty Moore Foundation (GBMF9461) and support from the US DOE under the Basic Energy Sciences programme (grant number DOE/BES DE-FG-02-05ER46200).

\end{document}